\documentstyle[preprint,aps]{revtex}
\def\P{\not\!\!\tilde{D}}
\def\A{\not\!\! A}
\def \Par{\not\!\partial}
\def\prd{Phys. Rev. D }
\def\npb{Nucl. Phys. B }
\def\prl{Phys. Rev. Lett. }
\def\plb{Phys. Lett. B }
\def\ann{Ann. Phys. (N.Y.) }

\preprint{SNUTP-96-068}
\begin{document}
%\baselineskip1.6\baselineskip
%\parindent1.0\parindent

\title{  The  effective action  of (2+1)-dimensional  QED:  the effect  of finite  fermion 
density } 
\author{Dae Kwan Kim\footnote{E-Mail:dkkim@power1.snu.ac.kr}
and Kwang-Sup Soh\footnote{E-Mail:kssoh@phyb.snu.ac.kr}}
\address{Department of Physics Education, 
Seoul National University,\\ Seoul 151-742, Korea}
\maketitle
\vspace{4cm}
\begin{abstract}
The effective action of (2+1)-dimensional QED with finite fermion density 
is calculated in a uniform electromagnetic field.
It is  shown that   the integer quantum  Hall  effect and  de  Haas-van  Alphen  like 
phenomena in 
condensed matter physics are derived directly from 
the effective action.
\end{abstract}
\draft
\pacs{11.10.Kk, 73.40.Hm} 
%\newpage
%\pagestyle{plain}
\section{introduction}
In (2+1)-dimensional spacetime, Fermi systems interacting through 
Maxwell 
field may have a dynamically induced  Chern-Simons (CS) term in 
the effective Lagrangian \cite{ja,ni}. That is, the low-energy 
effective action 
for the electromagnetic fields in the system, obtained by
integrating out the fermionic degrees of freedom,
has the induced CS  term as a parity-odd  part of the effective  
action. This CS term  could 
describe the quantum Hall effect (QHE) \cite{is}. 
In particular, an effective action at zero fermion density in uniform
electromagnetic field was obtained by Redlich, {\it et. al.} 
\cite{re,cl}, where the
coefficient of the 
induced CS term represents the Hall conductivity of the quantum Hall 
effect \cite{is}. However, real systems  in the experiment of QHE  
\cite{qhe} consist of finite 
density of electrons, therefore, it is
necessary to evaluate an effective action at finite fermion density 
and then to  examine the behavior  of the CS  term as the  external 
electromagnetic field  is 
varied. 

Recently, it was argued in ref.\cite{zi,sa} that the induced CS term 
in the presence of nonzero fermion density may describe 
the integer QHE; here, we use different method from  theirs, and 
calculate an effective action 
in a uniform electromagnetic
field at finite fermion density. Then we show directly that the 
coeffecient
of the induced CS term represents the QHE. Additionally, the free
energy of the system is obtained as a parity-even part of the effective
action. This exihibits a certain periodic dependence on the external 
field,
which is similar to the de Haas-van Alphen effect \cite{on}. 

The technique used here for deriving the effective action is the 
proper-time 
method, which was established by Schwinger first \cite{sh}, and 
generalized 
to the case of finite fermion density 
in $(3+1)$-dimensional spacetime by Chodos, {\it et. al.} 
several years ago \cite{ch}. 
  
In the section II, we calculate the effective action of 
(2+1)-dimensional QED 
with finite fermion  density in a  constant uniform electromagnetic  
field. From this  effective 
action, the integer quantum Hall effect and de Haas-van Alphen like 
phenomena in condensed 
matter physics are derived directly. In the last section, a summary of 
our results is given. 
\section{ EFFECTIVE ACTION  OF $(2+1)$-DIMENSIONAL QED WITH FINITE 
FERMION DENSITY}

We consider two-component Dirac fermions in a constant uniform 
electromagnetic field in (2+1)-dimensional spacetime.
The Lagrangian of the Fermi systems at a finite fermion density 
\cite{zi,shu} is given by
\begin{equation}
{\cal L} =\bar{\psi} (i \Par
-e \A)\psi-m\bar{\psi} \psi-\frac{1}{4}
F_{\mu \nu}F^{\mu \nu}-\mu \bar{\psi}\gamma^{0} \psi, 
\end{equation}
where $\mu$ is a chemical potential, $  \A= A_{\mu} \gamma^{\mu}$,
$F_{\mu\nu}=\partial_{\mu} A_{\nu}-\partial_{\nu} A_{\mu}$, and
 $\gamma^{0}=\sigma^{3}, \gamma^{1, 2}=i \sigma^{1, 2}$ with $\sigma^
{i}$ Pauli matrices.
Here, the last term in ${\cal L}$, $\mu \rho $ with $\rho= \bar{\psi} 
\gamma_{0}
\psi$, indicates that the system under consideration is composed of 
nonzero
density of fermions.
The fermion-mass term  in the  Lagrangian violates  parity $(P)$  and 
time-reversal  $(T)$ 
symmetries, and so
a $P$- and $T$-odd term, which  is the Chern-Simons term, is generated in 
the effective theory for the gauge field $F_{\mu\nu}$ \cite{ja}.

The one-loop contribution to the effective action $S_{eff}^{1} $ has 
the expression
\begin{equation}
S^{1}_{eff}= -i \mbox{Tr} \ln (i\Par-e \A-m-\mu \gamma_{0}),
\end{equation}
where $\mbox{Tr}$ denotes  the  summation over  spacetime  coordinates 
as well as spinor indices. 
This can be immediately rewritten as \cite{le}
\begin{eqnarray}
S^{1}_{eff}&=& -\frac{i}{2} \{ \mbox{Tr} \ln (\P-m)-\mbox{Tr} \ln 
(\P+m)\} \\ \nonumber
           & & -\frac{i}{2} \{ \mbox{Tr} \ln (\P-m)+\mbox{Tr} \ln 
(\P+m)\},
\end{eqnarray}
where $\tilde{D} ^{\mu}=(p^{0}-e A^{0}-\mu, p^{i}-e A^{i})$.
Note that the first  term in the  right hand side of Eq. (3), to 
be labelled as
$S_{eff} ^{odd}$, is explicitly odd under $m \rightarrow -m$ and
   corresponds to  the parity-odd CS action;   while,  the  second   
term, 
$S_{eff}^{even}$, is  the ordinary 
effective  action of even parity, which is  negative  value  of  
the free energy of the system. From the analysis of eigenvalues of
Dirac hamiltonian for the above Lagrangian Eq.(1), the transformation  $m 
\rightarrow -m$ results in the corresponding transformation of the
chemical potential $\mu$; $\mu \rightarrow - \mu$. This  property comes from the fact 
that 
in 2+1 dimensions, two types of mass parameter in  the Dirac equation are allowed, 
which holds only in 2+1 dimensions. So the induced
CS term must involve the terms of the form $m/ |m|$ or 
$\mu/|\mu|$ (see Eq.(17)). 

We first calculate the parity-odd part of the action $S_{eff} ^{odd}$.
To evaluate  the  $S_{eff} ^{odd}$ term, we may proceed in two steps 
\cite{ni,re}: first, the 
ground-state currrent density  of the  system, $\langle J_{\mu} 
\rangle$, is 
calculated, and then by integrating with respect to the vector 
potential $A_{
\mu}$, using the relation between an effective action and current 
density:
\begin{equation}
\frac{\delta S_{eff}}{\delta A^{\mu}}=\langle J_{\mu} \rangle,
\end{equation}
the parity-odd part of the effective action may be obtained. 
The ground-state  current $\langle  J_{\mu} \rangle$  may be  
expressed in   terms of  the Green's function $G$ 
\begin{equation}
\langle J_{\mu} \rangle = i e \mbox{Tr} (\gamma_{\mu}G),
\end{equation}
Here, the Green's function $G(x,y)$ is defined by
\begin{equation}
(i\Par-e \A-m-\mu\gamma_{0}) G(x,y)=\delta(x-y).
\end{equation}

We confine ourselves to the case where only
the magnetic field $B$ is turned on and the electric field $\vec{E}$ 
is zero.
At  the   end of   calculation,   we can   express   the evaluated   
effective   action in   a 
Lorentz-covariant form.
Under the Lorentz transformation, $F_{\mu \nu} \rightarrow F^{\prime}_{\mu 
\nu} 
$; therefore, the nonzero electric- as well as magnetic-field effects may be 
deduced.

As in ref.\cite{ch}, the solution of Eq.(6) for $G$ in the momentum 
space is given by
\begin{eqnarray}
G &=&(\P+m) \left\{-i\int_{0}^{\infty} d s \exp[is ( \P ^{2}-m^{2})]
\, \theta [(p_{0}-\mu) \mbox{sgn} p_{0}]  \right.  \nonumber \\
  & &+ \left. i\int_{0}^{\infty} d s \exp[-is(  \P^{2}-m^{2})] 
\, \theta[(\mu- p_{0}) \mbox{sgn} p_{0}] \right\}, 
\end{eqnarray}
where $\tilde{D}^{\mu}=(p^{0}-\mu, \vec{D})$ with $\vec{D}\equiv \vec{p}
-e \vec{A}$ and $F_{0 i}=0, F_{1 2}=-B$. 
Now we use the identity
\begin{equation}
\P^{2}=(p^{0}-\mu)^{2}- \vec{D}^{2}+ \frac{e}{2} \sigma_{\mu\nu}F^{\mu
\nu},
\mbox{  with } \sigma_{\mu \nu}\equiv -\frac{i}{2}[\gamma_{\mu},\gamma_
{\nu}].
\end{equation}

Inserting the expression for G, Eq.(7), into Eq.(5) and introducing
the proper-time parameter $s$ \cite{sh,ch} leads to
\begin{eqnarray}
\langle J_{\mu} \rangle &=&ie \, \mbox{tr}\, \gamma_{\mu}\langle x|-i
\int_{0}^{\infty} 
d s \exp[is (-m^{2}+(p^{0}-\mu)^{2}+ \frac{e}{2} \sigma \cdot F-
\vec{D}^{2})] \times \nonumber \\ 
                    & &(\P+m) \, \theta[(p^{0}-\mu) \mbox{sgn} p^{0}]  
                     +i\int_{0}^{\infty} d s \exp[-is (-m^{2}+(p^{0}-
\mu)^{2} \nonumber \\
                    & &+ \frac{e}{2} \sigma \cdot F-\vec{D}^{2})]\cdot 
                   (\P+m)\, \theta[(\mu-p^{0}) \mbox{sgn} p^{0}]|x 
\rangle , 
\end{eqnarray}
where $\langle x |$ denotes the eigenket of the spacetime coordinate 
operator $x^{\mu}$ in $\mbox{Tr}$ operation in Eq.(5).
Note that we are led  to a kind of  dynamical problem; that is,  $ 
\exp(-i s \vec{D}^{2})  $ play
the role  of an  evolution operator  in the  time variable  $s$, 
governed  by the  hamiltonian $\vec{D}^{2} $.  
Setting $| x \rangle \equiv | x^{0}\rangle |\vec{ x} \rangle$ and
$\langle  \vec{x}, s| \equiv  \langle \vec{x}| \exp(-i s 
\vec{D}^{2})  $,
$\langle J_{\mu} \rangle$ is rewritten as
\begin{eqnarray}
\langle J_{\mu} \rangle&=&ie \, \mbox{tr}\, \gamma_{\mu} \left\{ -i
\int_{0}^{\infty} 
d s \exp[-is( m^{2}- \frac{e}{2} \sigma\cdot F)] \int_{-\infty}^
{\infty}\frac{dp^{0}}{2
\pi} \exp[is(p^{0}-\mu)^{2}]\times \right. \nonumber \\
                   & &\langle \vec{x},s| D^{i}\gamma_{i} +(p^{0}-\mu)
\gamma_{0}
+m|\vec{x}\rangle  \, \theta[(p^{0}-\mu) \mbox{sgn} p^{0}]   
                   -i\int_{0}^{\infty} d s \exp[is( m^{2} -\frac{e}{2} 
\sigma\cdot F)] \times     \nonumber \\ 
                   & & \left. \int_{-\infty}^{\infty}\frac{dp^{0}}{2\pi
} \exp[-is(p^{0}-\mu)^{2}] \times  \langle \vec{x},s| D^{i}\gamma_{i} +
( p^{0}-\mu)\gamma_{0}+m|\vec{x}\rangle  \, \theta[(\mu-p^{0}) 
\mbox{sgn} p^{0}] \right\}. 
\end{eqnarray}

To evaluate the rather formal expression for $\langle J_{\mu}\rangle$, 
one may use the following quantities:
\begin{equation}
\mbox{tr} \; \exp(-\frac{i}{2}es\, \sigma _{\mu\nu} F^{\mu \nu}) 
= \cos(e s 
|B|)+i\, \frac{ \gamma^{\alpha} { }^{\ast} F_{\alpha}}{|B|}+ \sin(e s 
|B|), \nonumber \\ 
\end{equation}
and
\begin{eqnarray}
\langle  \vec{x}, s| D^{i}|\vec{x}\rangle            =&&0,   \\ 
\langle  \vec{x}, s| \vec{x}\rangle = \frac{-i}{4\pi} && 
\frac{e |B|}{ 
\sin (e s |B|)}, \nonumber 
\end{eqnarray}
where ${ }^{\ast} F_{\alpha}= \frac{1}{2} \epsilon_{\alpha \beta \gamma} 
F^{\beta \gamma}$ with ${ }^{\ast} F_{0}=-B,{ }^{\ast} F_{i}=0$. 

Note that only the time component of $\langle J_{\mu} \rangle $, 
$\langle J_{0} \rangle $, is sufficient in determining the
coefficient of CS term \cite{ni,re}.
Substituting Eq. (11) and (12) into Eq. (10) and using $\int_
{-\infty}^ {\infty}d x 
\exp(i s x^{2})= \sqrt{\pi/s} \exp( i \pi/4)$, one 
may get that
\begin{eqnarray}
\langle J_{0} \rangle&=& \frac{me^{2}}{4 \pi^{3/2}}\exp(\frac{i\pi}{4})
{ }^{\ast} F_{0}
\int_{0}^{\infty} \frac{d s}{\sqrt{s}} \exp(-is m^{2})-\frac{me^{2}}{2 
\pi^
{2}} { }^{\ast} F_{0} \; \mbox{Re} \int_{0}^{\mu}dx \int_{0}^{\infty} 
d s \exp[(is (x^{2}-
\nonumber \\              & &m^{2})]  
            -\frac{e^{2}}{2  \pi^{2}}   { }^{\ast}   F_{0}\;   
\mbox{Im} \int_{0}^{\mu}dx\,   x 
\int_{0}^{\infty} 
d s \exp[(is (x^{2}-m^{2})] \cot( e s |B|),
\end{eqnarray}
where $\mbox{Re}$ $(\mbox{Im})$ denote the real (imaginary) part of 
the corresponding quantity.

To obtain the final expression for $\langle J_{0} \rangle$ in Eq.(13), 
one has to take the 
integration over both $s$ and $x$ variables. The integration over $s$ 
in the
first term is taken by letting $s \rightarrow -i\, s$. 
As may be checked, in  the calculation of the second term in the 
r.h.s. of Eq. (13), one had better integrate over $s$ variable first and
then over $x$ variable. In this way, one get the $\, \theta (\mu^{2}-
m^{2})$ function.
On the other hand, some care is needed in the calculation of the
third term  in  the above  equation;  in this  case,  an integration  
over  $x$ may  first  be 
performed. And then in the integration over $s$, the integration
contour in the real axis $(0,\infty)$ does have the meaning of $(0-i 
\epsilon, \infty-i \epsilon)$ \cite{ch,el}. 
Thus, the third term in $\langle J_{0}\rangle $ in Eq. (13) leads to 
the following expression:
\begin{equation}
\varepsilon (\mu) \, \theta(\mu^{2}-m^{2})\, \pi^{2} \left\{\frac
{\mu^{2}-m^{2}}{2 e |B|
}+\sum_{n=1}^{\infty}\frac{1}{ \pi n}\sin \left[ \pi n \, \frac
{(\mu^{2}-m^{2})}{e |B|}\right]
 \right\},
\end{equation}
where $ \varepsilon(\mu)$ is defined  by $ \varepsilon(\mu)=\mu/|
\mu|$ and  $\theta(x) $ is 
the step function such that
$\theta(x)=1 $ for $x>0 $ and $\theta(x)=0 $ for $x<0 $.
Adding up three terms in Eq.(13), we have a more familiar expression
for $\langle J_{0}\rangle$: 
\begin{eqnarray}
\langle J_{0} \rangle &=& \frac{e^{2}}{4\pi}
\frac{m}{|m|}
[(1-\theta(\mu^{2}-    m^{2})]{ }^{\ast}   F_{0}-\frac{e^{2}}{2\pi}
\varepsilon    (\mu)\,    \,   
\theta 
(\mu^{2}-m^{2})
\left\{\frac{\mu^{2}-m^{2}}{2 e |B|}\right. \nonumber \\
                      & &\left.   +\sum_{n=1}^{\infty}\frac{1}{ \pi n}
\sin \left[ \pi n \, \frac{ (\mu^{2}-m^{2})}{e |B|} \right]
         \right\} { }^{\ast} F_{0}.
\end{eqnarray}
Here, the charge  density  vector  $\langle J_{0}  \rangle$  is one  component  of the  
covarint 
current  vector  $\langle   J_{\mu} \rangle$.   Note  that   under a Lorentz 
transformation, $ |B|= |{  }^{\ast} F|$ is  an invariant  quantity;  that is, $  |{ }^{\ast} 
F|=|B| 
\rightarrow |{ }^{\ast} F^{\prime}|=\sqrt{{B^{\prime}}^{2}-{E^{\prime}}^{2}} $.
The present situation is the zero temperature limit of relativistic thermdynamics.
Thermodynamics can be formulated in the relativistically covariant way \cite{wl}. 
The chemical potential $\mu$ is a Lorentz scalar and defined by the  value in the  rest 
frame 
of Fermi gases; therefore, $\mu^{2} \rightarrow \mu^{2}$ under the Lorentz 
transformation. 
One may write the parity-odd part of the effective action in the covariant form 
\begin{equation}
S^{odd}_{eff}=
\frac{1}{2}\int \langle J_{\mu} ^{\prime}\rangle A^{\prime}{ }^{\mu} 
d^{3}x^{\prime}. 
\end{equation}
The corresponding effective Lagrangian is as follows:
\begin{eqnarray}
{\cal L}^{odd}_{eff}&=&   \frac{e^{2}}{16\pi}   \left\{  \frac{m}
{|m|}    [(1-\theta(\mu^{2}- 
m^{2})]-2\, 
\varepsilon (\mu)\,  \, \theta(\mu^{2}-m^{2}) \left[  \frac{\mu^{2}-
m^{2}}{2 e  | { }^{\ast} F|} 
\right. 
\right. \nonumber \\
             & &  \left. \left.  +\sum_{n=1}^{\infty}\frac{1}{\pi  n}
\sin \left(  \pi n   \, \frac{ 
(\mu^{2}-m^{2})}{e |{ }^{\ast}
  F|} \right)  \right] \right\} \epsilon^{\alpha \beta \gamma}
A_{\alpha}F_{\beta \gamma},
\end{eqnarray}
where the expression has the covariant form and the prime was suppressed
for the simplicity  of notation. Note carefully that taking the derivative 
of Eq.(17) with respect to $A^{0}$ leads to Eq.(15). The  first $\mu$-independent  term 
was the  obtained one  in ref.\cite{re,cl}. 
It is interesting to notice that the CS term for a vacuum fermion 
system (without the $\mu$-dependent term  in the Lagrangian) vanishes 
by  the presence of arbitrary small fermion density. One may see that
the $\varepsilon (\mu)$ term is odd under $m \rightarrow -m$; that is, this 
transformation leads to the corresponding transformation in $\mu$ 
such as $\mu \rightarrow -\mu$.

The coefficient of  CS term corresponds to the Hall conductivity 
\cite{is}.
This may be easily calculated in the frame where the electric field 
$\vec{E}$ vanishes. Then, the chemical potential $\mu$ satisfies  
$\mu^{2}= m^{2}+               2|e 
B|n $, when the fermions in the system occupy Landau levels up to 
certain integer $n$.   
And the infinite series term \cite{el} in ${\cal L}^{odd}_{eff}$ 
vanishes. Then, Eq. (17) is reduced as follows:
\begin{eqnarray}
{\cal L}^{odd}_{eff}&=&- \frac{e^{2}}{8\pi}\varepsilon(\mu)\,
\frac{\mu^{2}-m^{2}}{2 e | { }^{\ast} F|}
\epsilon^{\alpha \beta\gamma} A_{\alpha}F_{\beta\gamma} \\
             &=&  - \frac{e^{2}}{8\pi}\varepsilon(\mu)\, n \,
\epsilon^{\alpha \beta\gamma} A_{\alpha}F_{\beta\gamma}. \nonumber
\end{eqnarray}
Thus we get the Hall conductivity $\sigma _{xy}$: 
\begin{equation}
\sigma_{xy}= \frac{e^{2}}{2\pi} \varepsilon(\mu)
 \; n,
\end{equation}
where $n$ is some integer number corresponding to the filled Landau 
levels in   the system.  In  this  case   of a    finite fremion   density \cite{rj},  the   
conductivity is 2 
times  the 
one for the vacuum system for each state $n$.
This is the well known integer Quantum Hall effect.

We now calculate the parity-even part of the effective action, 
$S_{eff}^{even}$, in Eq.(3). 
This may be calculated as in the $(3+1)$-dimensional case,
\begin{eqnarray}
S^{even}_{eff}&=&  -\frac{i}{2}   \{  \mbox{Tr}   \ln   (\P-m)+\mbox
{Tr}  \ln  (\P+m)\}    \\ \nonumber
              &=& -\frac{i}{2}  \mbox{Tr} \ln (\P ^{2}-m^{2}).
\end{eqnarray}

As the odd part, we first consider pure magnetic case and then
extend to the general case at the end of the calculation.
Following the procedure in ref.\cite{sh,ch} and using the Green's 
function in Eq. (7), one may obtain $S_{eff}^{even}$: 
\begin{eqnarray} 
S^{even}_{eff}&=&\frac{i}{2} \mbox{tr}  \int_{0}^{\infty} \frac{d s}{s}
\langle x| \exp[ is ({\P}^{2}-m^{2})]
\, \theta[(p_{0}-\mu)\, \mbox{sgn} p_{0}]|x \rangle \\ \nonumber 
              & &+\frac{i}{2} \mbox{tr}  \int_{0}^{\infty} \frac{d s}{s}
\langle x| \exp[ -is ({ \P}^{2}-m^{2})]
\, \theta[(\mu-p_{0}) \, \mbox{sgn} p_{0}]|x \rangle . 
\end{eqnarray}
Substituting Eq.(8) into Eq.(21) and using Eq.(11) and (12), we have 
the following expression for the
parity-even effective Lagrangian ${\cal L}_{eff}^{even}$:
\begin{eqnarray}
{\cal L}_{eff}^{even}&=& - \frac{\exp(i\pi/4  )}{8 \pi^{2/3}}\int_{0}^
{\infty}  \frac{d   s}{s^{5/2}} \exp 
(-is m^{2})e s |B|\cot (e s |B|) \\ \nonumber
              &   &-\frac{1}{4\pi^{2}}\mbox{Re}  \int_{0}^{\mu}dx   
\int_{0}^{\infty}  \frac{d 
s}{s^{2}} \exp[is (x^{2}-m^{2})] e s |B|\cot (e s |B|) .
\end{eqnarray}

The first term in  the r.h.s. of Eq.(22) is calculated by deforming 
the path of integration $s \rightarrow -is$; for $m=0$ it can be 
analytically integrated
\cite{re}.
The integration over the $s$-variable in the second term is performed 
by choosing a  proper  contour  in   the complex   $s$-plane just   
as  the third   term in    Eq.(13) \cite{ch}. Then, it leads to the 
following two terms:
\begin{eqnarray}
  &  &\frac{  e   |B|}{2\pi^{2}}\int_{0}^{\mu}dx \;   \mbox{Im}  \sum_
{n=1}^{\infty}\frac{1}{n 
}\exp\left[i\pi n\frac{x^{2}-m^{2}}{e |B|}\right]\, \theta(x^{2}- 
m^{2}) \nonumber \\
  &  &+ \frac{1}{12\pi}\, \theta(\mu^{2}- m^{2})(|\mu|-m)^{2}(|\mu|+2m).
\end{eqnarray}
Thus, expressing this in the covariant form as in Eq.(16), we get the 
parity even part of the one-loop effective action
\begin{eqnarray}
{\cal L}_{eff}^{even}&=&    \frac{1}{8  \pi^{2/3}}\int_{0}^{\infty}  
\frac{d   s}{s^{5/2}} \exp    (- m^{2} s) 
[ e s |{ }^{\ast} F| \coth (e s |{ }^{\ast} F|)-1 ]   \nonumber \\
              & &+\frac{1}{12\pi}\, \theta(\mu^{2}- m^{2})(|\mu|-m)
^{2}(|\mu|+2m) \nonumber \\
              & &+\frac{(e|{ }^{\ast}   F|)^{3/2}}{\pi^{2}}  
\sum_{n=1}^{\infty}\frac{1}{n^{3/2}  }\sin 
^{3/2}        \left(\pi       n\frac{\mu^{2}-m^{2}}{2e        
 |{ }^{\ast}         F|}\right)\cos^{1/2}\left(\pi 
n\frac{\mu^{2}-m^{2}}{2e |{ }^{\ast} F|}\right),
\end{eqnarray}
where for renormalization one term has been added to the first term on
the right-hand side.
Notice that the first $\mu$-independent term in this effecftive 
Lagrangian is the 
previously obtained result in ref.\cite{re}. 
The third term in the  r.h.s. of Eq.(24)  shows  a periodic behavior 
as  $ |{ }^{\ast} F|$  is 
decreased; 
the analogous term  also appears  in the  effective Lagrangian  of 
$(3+1)$-dimensional  QED 
\cite{ch,el}. In ref.\cite{el} it is argued   that the infinite series 
term  when $\vec{E}=0$ may 
describe the de Haas-van Alphen effect \cite{on} in condensed matter 
physics. 
\section{ DISCUSSION}

In this article we obtained an exact one-loop effective action of 
(2+1)-dimensional QED with  finite fermion density  in a  uniform 
electromagnetic field.  The 
obtained total effective Lagrangian ${\cal L}^{1}_{eff}={\cal L}^{odd}_
{eff}+{\cal L}^{even}_{eff}
 $ in Eq.(17) and Eq.(24) agrees
with the previously calculated one in ref.\cite{re,cl} where the 
chemical
potential $\mu$ dependent term is not introduced into the Lagrangian. 

We showed that the coefficient of the induced CS term exhibits the 
step-function 
behavior, when the fermions fill some Landau levels completely, 
and that this corresponds to the phenomena of the integer quantum 
Hall effect  \cite{zi,sa}.  In particular,   the induced fermion   density  in  the vacuum 
vanishes if a 
finite density of fermions is added 
to the  system. Then,  the Chern-Simons  term corresponding  to the  induced fermion 
density in 
the vacuum disappears. Similar  behavior has already   been observed in  ref. \cite{da}; 
that 
is, the  fermion density   in the vacuum   corresponding to the   same  system under 
consideration 
in this  paper  evaporates   for any  nonzero  temperature   $T \neq   0$.   These 
nonanalytic 
behavior under 
the effect  of finite  temperature or   finite chemical   potential seems to  be  a  very 
general 
phenomenon.

In  addition, the third term in ${\cal L}_{eff}^{even}$ in Eq. (24) 
for $\vec{E}=0$ 
has a periodic  behavior as  a function  of $(\mu^{2}-m^{2})/2e  |B|$. 
This  agrees with the 
frequency for the de Haas-van Alphen effect \cite{el}. 
So the infinite 
series over $n$ in Eq.(24) seems to be related to the de 
Haas-van Alphen effect \cite{on}. 

\vspace{0.5cm}
\noindent
%\acknowledgements
{\large{\bf Acknowledgements}}
\vspace{0.5cm} \\

This work was supported by the Korea Science and Engineering Foundation 
(KOSEF), the Center  for Theoretical Physics(SNU),  and  the Basic  Science Research 
Institute 
Program,  Ministry of Education,
 1996, Project No. BSRI-96-2418.

\end{document}